# Impact of 150keV and 590keV proton irradiation on monolayer MoS2

Burcu Ozden[1], Ethan Khan[2], Sunil Uprety[3], Tianyi Zhang[2], Joseph Razon[1], Ke Wang[4], Tamara Isaacs-Smith[3], Minseo Park[3], Mauricio Terrones[2]

[1]Engineering and Science Division, Penn State Abington, Abington, PA, 19001, USA
[2] Department of Materials Science, Penn State, University Park, PA 16802, USA
[3] Department of Physics, Auburn University, Auburn, Alabama 36849, USA

## Abstract

We present a comprehensive study on the effects of proton irradiation at different energies (150 and 590 keV) with the fluence of 1x $10^{12}$ proton/$cm^2$ on monolayer $MoS_2$. This study not only improves our understanding of the influence of high-energy proton beams on $MoS_2$ but also has implications for radiation-induced changes in device processing and engineering of devices from multilayer $MoS_2$ starting material. Increasing defect density with decreasing proton irradiation energy was observed from photoluminescence spectroscopy study. These defects are attributed to sulfur vacancies observed through x-ray photoelectron spectroscopy analysis and confirmed by transmission electron microscope imaging. Scanning electron microscopy images showed the creation of grain boundaries after proton irradiation. A higher degree of surface deformation was detected with lower irradiation energies through atomic force microscopy. Inter-defect distance is increased with the increase in proton energy irradiation as estimated by transmission electron microscopy imaging. Raman spectroscopy reveals negligible structural changes in the crystal quality after the irradiation. These deformation damages due to proton irradiation are insignificant at the $MoS_2$ layer. Based on the overall influence of low energy proton irradiation on the material characteristics, ML-$MoS_2$ materials can be considered robust and reliable building blocks for 2D material based devices for space applications.

## Introduction

We live in an era where space tourism finally becomes a reality. Space tourism is expected to become a large industry within the next decade as more private sector companies promote it[1]. This ambitious goal can only be achieved through (1) the discovery of radiation-hard materials that complement the existing and emerging semiconductor materials in the extreme space environment and (2) the development of radiation-robust technologies specifically designed to deliver targeted device functionalities such as high-performance, low-power, and cost-effective.

In this regard, two-dimensional materials (2DMs) have attracted considerable attention for space applications due to potential use as a replacement of traditional semiconductors for next-generation electronic devices and circuits. Among these materials, molybdenum disulfide ($MoS_2$) is a promising candidate due to its attractive semiconducting properties, including high mobility[2], strong light emission[3,4] and absorption[5], bandgap tunability[5], and remarkable mechanical properties[6]. However, not much is known about how the particle radiation affects these materials even though they show great potential for space-bound electronics.

Studying the effects of proton irradiation on MoS$_2$ can help us understand the response of optoelectronic devices fabricated from these materials operating in hostile radiation environments such as space radiation environment, high-altitude flights, and military aircraft, satellites, nuclear reactors, and particle accelerators. Radiation-induced displacement damage can negatively impact the device performance of 2D materials such as MoS$_2$. These damages cause a significant reduction in the minority carrier lifetime of the optoelectronic devices made by these materials[7,8,9]. Besides, studying the effect of radiation is essential to discover new processing methods for materials and devices since energetic particles are commonly used for plasma surface treatments[10]. Furthermore, by evaluating radiation effects in the 2D material MoS$_2$, we also understand the new ways of engineering material properties such as defect concentration[11], ferromagnetism[12], and superconductivity[13] through external treatments.

The effect of proton irradiation on the electronic properties of 2D materials systems has been studied previously.[8-10,12-18] Researchers[12,14,15] used the proton irradiation to induce the weak ferromagnetism on the pristine MoS$_2$ with the improved transport property. Kim *et al.*[16] showed that the increase of conductance on backgated MoS$_2$ field-effect transistors (FETs) due to high energy (10 MeV) proton irradiation-induced traps. High responsivity ( ~1.04 A/W) and photo gain of tri-layer MoS$_2$ metal-semiconductor-metal photodetectors under 2 MeV proton illumination radiation were shown by Tsai *et al.*[17] In a study done by Arnold *et al.*[1], the back-gated MoS$_2$ FETs maintained high on currents and on/off ratios even after 2 MeV proton irradiation. The electronic properties of 1.8 MeV proton irradiated vertically stacked MoS$_2$/interlayer/MoS$_2$ heterostructures studied by Wang *et al.*[18] They did not observe any significant changes after irradiation.

While proton radiation effects on the electrical properties of MoS$_2$ studied extensively, less is known about the optical, chemical, and morphological characteristics of MoS$_2$. Furthermore, most of the existing studies conducted for higher energies of proton irradiation. In a proton irradiation study of MoS$_2$/polyimide at 25 keV with the flux of 2.5 × 10$^{14}$ cm$^2$/s$^{-1}$, Liu *et al.*[19] found that proton irradiation can break the polymer bond easily and then form the graphite-like structure at the surface of the samples. Wang *et al.*[9] demonstrate enhanced photoluminescence and direct bandgap emission in 100 keV proton irradiated MoS$_2$ flakes. In the work of Tongay *et al.*[20] defect activated photoluminescence and a blue shift in the direct transition PL peak after the α-particle of monolayer (ML) MoS$_2$. Higher energy (10 MeV) proton irradiation studies of monolayer ML-MoS$_2$ focused on the correlation of the surface potential revealed that the Fermi-level of the ML-MoS$_2$ shifted upwards after proton irradiation[11]. There has been only one study by Foran *et al.*[21] with photoluminescence (PL) spectroscopy and transmission electron microscopy (TEM) imagery data to enhance our understanding of the fundamental physical processes affecting material properties. In this study, they observed defect-activated PL and a blue shift in the direct transition PL peak of MoS$_2$ flakes on the substrate after proton irradiated at 100 keV with fluences ranging between 10$^{13}$ and 10$^{16}$ cm$^2$/s$^{-1}$.

Nevertheless, it is not fully understood how radiation at different energies would affect the optical, chemical, and morphological properties of ML-MoS$_2$. It is of great importance to investigate the effect of irradiation on the 2DMs if they are used in radiation-hard electronics. We present a comprehensive study on the effects of proton radiation effects on ML-$MoS_2$ by rivalling a particular space radiation environment using relatively low energy (150 keV) and (590keV) proton beam with the fluence of 1×10$^{12}$ protons/cm$^2$. Studying the effects of proton irradiation for space applications is specifically essential because of the appearance of energetic protons in cosmic radiation. According to modeling radiation levels in the thermosphere using the SPace ENVironment Information System (SPENVIS) by Vogl *et al.*[22], the energy spectrum for protons in low Earth orbit (LEO) ranges from 100 keV to 400 MeV. The vast majority of spacecraft, antennas, pointing mechanisms are directly exposed in the LEO space environment. Even though the trapped particle belts in LEO essential for life on Earth, heavier ions like protons can cause great danger to any spacecraft orbiting through these particle belts since they experience higher stopping power. Low-energy particles are disadvantageous because only these particles can deposit significant amounts of energy to the spacecraft walls and act as a non-ideal high-pass[22]. Besides, Mathew *et al.*[23] showed that lower proton energies (200, 500, 1000, and 2500 keV) caused higher the potential displacement damage on graphene samples due to the higher stopping power of the 2D materials at lower energies.

Therefore, this paper concentrates upon understanding the effects of low energy proton irradiation on $MoS_2$. The pristine and irradiated ML-$MoS_2$ samples were characterized by comparing the phenomenological changes in the optical, chemical, and morphological properties to enhance our understanding of the fundamental physical processes affecting material properties. The crystal quality was examined with Raman spectroscopy. The possible effects on defects/ traps were probed using photoluminescence (PL) spectroscopy. The surface compositional analysis was performed via x-ray photoelectron spectroscopy (XPS), and surface morphology was inspected using atomic force microscopy (AFM), scanning electron microscopy (SEM), and transmission electron microscopy (TEM).

In this article, we showed the effects of 150 keV protons on the $MoS_2$ samples are more severe compared to 590 keV protons. However, overall quantified changes were considered insubstantial. Based on the observed level of influence of low proton energies on material characteristics, the $MoS_2$ has high endurance for exposure to the low-energy proton beam. Hence, we conclude that proton irradiation is no concern for $MoS_2$ even for lower proton energies in LEO. Our findings highlight the importance of evaluating the impact of damage to the optical, optical, morphological, and chemical properties of these materials if used for space-bound electronics.

**Experiments**

The ML $MoS_2$ films with 1.2 nm thickness were grown on $SiO_2$ (300nm)/Si$^{++}$(500µm) substrates using a chemical vapor deposition (CVD) method. In the two steps growth, NaBr particles were first deposited on the substrate through thermal evaporation. Metal dioxide (e.g., $MoO_2$) was used as precursors, which were reacted with S vapor at 770 °C. Details of the synthesis can be found elsewhere[24]. Optical images of the pristine and proton irradiated sample shown in Fig. 1a, b.

The Atomic Force Microscopy (AFM) was performed on a Bruker Dimension Icon (Bruker Nano Inc, Santa Barbara). Topography was collected in PeakForce Tapping mode using a Scanasyst Air probe (0.5 N/m spring constant, 2nm nominal tip radius) and a 0.5-1.5 nN setpoint. The thickness is 1.2 ± 0.1 nm, determined from the AFM height profile (Fig. 1c), suggesting that the flake is monolayer even after irradiation.

Two pieces were irradiated at room temperature with 150 keV (Sample 1) and 590keV (Sample 2) proton beams of $1\times10^{12}$ protons/cm$^2$ fluences generated from a 2MV dual-source Tandem Pelletron accelerator located at Auburn University. The accuracy of the proton beam energy (monitored by the beam current) was maintained at 99.9%.

The Raman and PL spectra were collected at room temperature in ambient conditions on pristine and irradiated samples. The Renishaw inVia Raman microscope system with 1800 grooves/mm grating and a 532 nm laser at 1 mW laser power as excitation source was used to study the optical properties of the 1L-MoS$_2$ using Raman and PL spectroscopy.

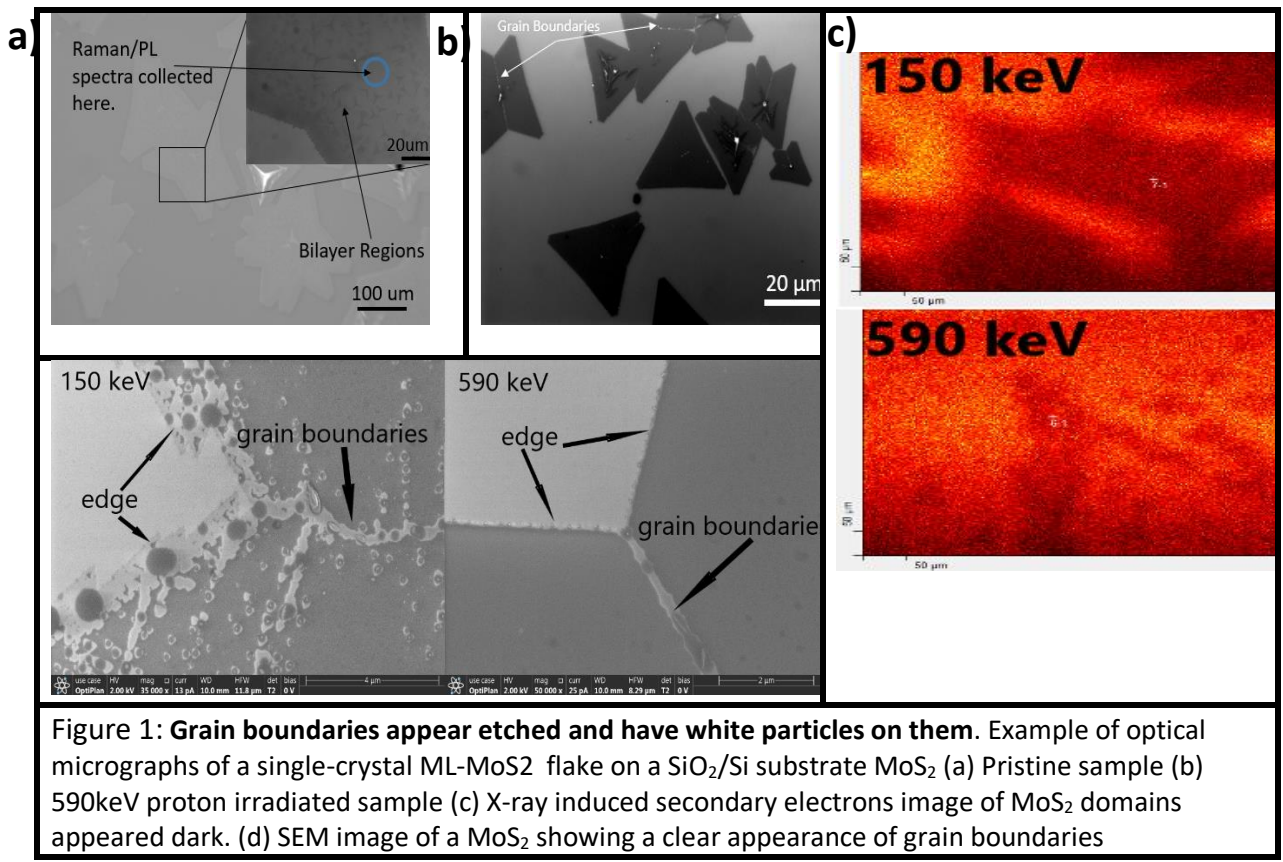

Figure 1: **Grain boundaries appear etched and have white particles on them**. Example of optical micrographs of a single-crystal ML-MoS2 flake on a SiO$_2$/Si substrate MoS$_2$ (a) Pristine sample (b) 590keV proton irradiated sample (c) X-ray induced secondary electrons image of MoS$_2$ domains appeared dark. (d) SEM image of a MoS$_2$ showing a clear appearance of grain boundaries

X-ray photoelectron spectroscopy (XPS) experiments were performed using a Physical Electronics Versa Probe II instrument equipped with a monochromatic AlK$_\alpha$ x-ray source (hv = 1,486.7 eV) and a concentric hemispherical analyzer. Charge neutralization was performed using both low energy electrons (<5 eV) and argon ions. The binding energy axis was calibrated using sputter cleaned Cu (Cu $2p_{3/2}$ = 932.62 eV, Cu $3p_{3/2}$ = 75.1 eV) and Au foils (Au $4f_{7/2}$ = 83.96 eV)[25]. Peaks were charge referenced to the CHx band in the carbon 1s spectra at 284.8 eV. Measurements were made at a take-off angle of 45° with respect to the sample surface plane. This resulted in a typical sampling depth of 3-6 nm (95% of the signal originated from this depth or shallower). Quantification was done using instrumental relative sensitivity factors that account for the x-ray cross-section and inelastic mean free path of the electrons.

In-plane transmission electron microscope (TEM) images were collected on an FEI Titan3 G2 equipped with double correctors. We tried to minimize beam-induced lattice damage by taking images with a beam at 80 keV, producing a probe current of 35 pA. Furthermore, all the images were taken after all the other measurements were completed. We have not observed any damage accumulation after multiple imaging of the $MoS_2$ flakes.

SEM Imaging was performed using a Thermo Fisher Scientific Apreos, Field-Emission Scanning Electron Microscope under the high-vacuum ($10^{-6}$ Torr) with the beam current 13pA - 1nA.

## Results and Discussion

Characteristic Photoluminescence (PL) (Fig. 2a) and Raman (Fig. 2b) spectra confirm the monolayer nature of the pristine $MoS_2$ crystals. The PL spectrum shows a dominant emission band from A exciton at ~1.84 eV, with an FWHM (full-width at half maximum) of ~50 meV and a weak B exciton band, typical for high-quality ML-$MoS_2$ single crystals for the pristine sample. After irradiation, A peak slightly red-shifted to ~1.82 eV for both 150 keV and 590 keV proton irradiation respectively from the primary direct band emission at 1.84 eV. Similar behavior has been observed for monolayer and bilayer $MoS_2$ after 100 keV proton irradiation by other researchers[9,21]. These slight changes in electronic bandgap would alter the electronic device performance, perhaps by changing doping levels, metal-contact work function alignment, and ON/OFF ratios if they were used in radiation hard environments[21]. On the other hand, we did not detect additional defect-activated photoluminescence. Based on the line broadening and intensity quenching of the PL spectrum, we can say lower energy proton irradiation causes higher defect creation in ML-$MoS_2$. Possible S vacancies observed through XPS and TEM analysis confirms this result (Fig. 3 and 5). The PL intensity quenched by a factor of 7.5x and 1.7x for 150keV and 590keV due to irradiation-induced defects. These defects might cause non-radiative recombination and strongly affect the performance of bipolar devices such as BJTs and diodes.[21] The quenching of the spectrum can also be attributed to the progressive ablation of the material caused by the ions bombarding the $MoS_2$ flakes[26]. We rule out the effect of room temperature defect annealing on the change in intensity of the PL peaks since the spectrum remains the same after all the measurements completed. Owing to the fact that the $MoS_2$ films are already a monolayer, and exhibiting a direct transition, it would be expected the intensity to decrease after exposure to radiation[21]. The FWHM values of proton 150 keV and 590 keV $MoS_2$ flakes were broadened 21% and 9%, respectively. These behaviors are likely due to the creation of defects, bound excitons, and recombination centers causing the non-radiative recombination at the $MoS_2$-$SiO_2$ interface and shorten the lifetime of the photoexcited carriers[27]. If these observed behaviors are eminent, proton irradiation at given energies can affect the performance of devices fabricated from these materials. For example, Kim *et al.*[16] show that the high-energy (10MeV) proton irradiation of MoS$_2$ FET can induce trap states at interfaces and within the gate oxide, degrading the transistor's performance. On the other hand, Foran *et al.*[21] showed the recovery of defect-associated linewidth broadening of the direct transition after annealing. Therefore, it might be possible to recover indicates device degradation due to radiation-induced defects detected by PL. Observed changes in PL intensities due to proton irradiation of MoS$_2$ is relatively negligible compared to optoelectronic materials found in the literature where PL intensity reduced ~100x to ~10000x.[21]

PL data itself is not enough to make a quantitative analysis of the defect concentration and to say whether these produced defects would affect the device performances. Therefore, Raman measurements performed to quantify the defects. Nevertheless, Raman measurements did not reveal any significant contribution of the defects that would result in a change in the electronic band structure of MoS$_2$. Furthermore, as suggested by Foran *et al.*[21], time-resolved PL studies of pristine might be useful to analyze the PL intensity reduction since the displacement damage is strongly dependent on the initial starting material quality and minority carrier lifetime considering the TEM studies did not identify any atomic defects in pristine samples[24].

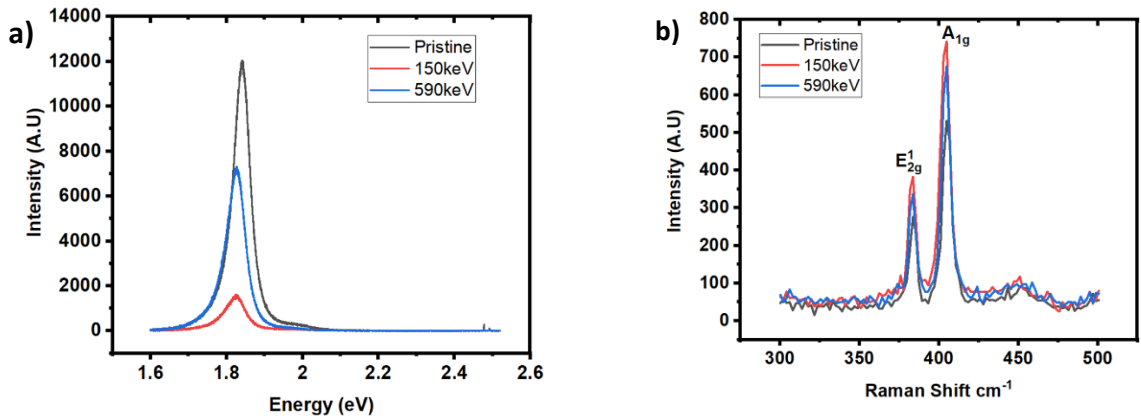

Figure 2: **Optical properties of MoS$_2$ monolayers.** (a, b) PL and Raman spectra acquired from single-crystal pristine and irradiated MoS$_2$ monolayers at room temperature with a 532 nm laser excitation. Both spectrums maintained the main shape with the change in intensities after irradiation.

Raman spectroscopy has proven to be a useful tool to determine, not only the number of layers[28] of MoS$_2$, but also the built-in strain[29] in the layers, their doping level[30], and the defects controllably created in ML MoS$_2$ by calculation the inter-defect distance (LD)[31]. Figure 2 shows a comparison of the Raman spectra measured for pristine and proton irradiated ML-MoS$_2$ samples at different energies of 150keV and 590keV. In large crystals >60 μm, there is an inhomogeneous strain between centers and edges. Furthermore, lower Raman intensity near the edges of the irradiated samples was observed, indicating higher defect density near edges in irradiated samples. Raman spectra are compared between the center points of ~20 μm triangles to avoid the complexity. The Gauss-Lorentz function is used to fit the Raman spectrum. Two vibrational modes dominate the mom-resonant Raman spectrum of MoS$_2$: the in-plane ($E^1_{2g}$), which is due to vibrations of two S atoms with respect to the Mo atom and out-of-plane ($A_{1g}$) which corresponds to the vibrations of S atoms perpendicular to the layer. Across the pristine sample, observed peak frequencies are uniform, indicating that $E^1_{2g}$ and $A_{1g}$ peaks can be used to identify the number of layers of an ultrathin $MoS_2$ flake with much higher accuracy[28,32]. There are two characteristics peaks in the Raman spectrum of pristine $MoS_2$ sample, the $E^1_{2g}$ (centered at ~ 383.6cm$^{-1}$) and $A_{1g}$ (centered at ~ 404.8cm$^{-1}$). The peak frequency difference (Δk) between $E^1_{2g}$ and $A_{1g}$ 21.2cm$^{-1}$ and the ratio $A_{1g}/E^1_{2g}$ =1.055 ≅ 1, which can indicate the monolayer-$MoS_2$ [28]. Our Raman results were consistent with the Raman results of $MoS_2$ reported from other groups[11,33]. Well-defined $E^1_{2g}$ and $A_{1g}$ modes appeared after irradiation. It is known that Δk values increase from monolayer to bulk.[34] In this study, we did not observe any changes in the values of Δk after irradiation. The position of the $E^1_{2g}$ and $A_{1g}$ peaks seems rather insensitive to proton irradiation at low energies since both Raman peaks show approximately 0.4 cm$^{-1}$ variations (redshift) between the pristine and irradiated samples with 590 keV irradiation. The redshift $E^1_{2g}$ can be attributed to the weakening of the Mo–S bonds involved in the in-plane vibration increasing vacancy concentration. The ion-induced defects in 2D materials are usually associated with doping[35] and strain[36], which can affect the vibrational modes. Previously, doping effects have been observed with the blue shift of $A_{1g}$ while the peak position of the $E^1_{2g}$ peaks remains unaffected[37]. However, these results are inconsistent with our observation; hence, it is less likely to have the doping effects here. The shift in peaks position can also be associated with an induced strain localized around defects. While the decrease in the peak positions associated with in-plain tensile strain, the compressive strain is attributed to the increases in the peak positions[36,38]. In our case, we observed a decrease in the $E^1_{2g}$ and $A_{1g}$ peak positions, which can be attributed to tensile strain localized around sulfur vacancies observed in morphological images. Previously, Rice *et al.*[29] observed a shift in the $E^1_{2g}$ and $A_{1g}$ modes of 2.1 and 0.4 cm$^{-1}$ per % of uniaxial strain, respectively. Hui *et al.* [39] measured that the applied biaxial strain of % $E^1_{2g}$ and $A_{1g}$ modes shifted by ~4.7 cm$^{-1}$ and ~3 cm$^{-1}$, respectively. In this particular case of 590 keV proton irradiation, we can then estimate a maximum strain level of ~0.2% in uniaxial strain and ~0.09% in biaxial strain. There is no change in the Raman peaks with 150 keV proton irradiation. This lack of significant structural changes in the $MoS_2$ Raman modes indicates insignificant damage to the crystal structure of $MoS_2$ lattice. Thus, we can conclude that proton irradiation at 150 keV and 590 keV with the fluence of 1×10$^{12}$ protons/cm$^2$ does not significantly strain our samples.

The FWHM of the $E^1_{2g}$ peak increased by 1% and 4%, respectively, for 150 keV and 590 keV proton irradiation. While the FWHM values of $A_{1g}$ modes increased 4% for 150 keV proton irradiation, it decreased 5% for 590 keV proton irradiation. A minor change of FWHM values indicate that the lattice structure has been preserved in the near-surface region of the irradiated sample[12]

The intensity of $E^1_{2g}$ and $A_{1g}$ modes is slightly modulated by the energies of proton irradiation but maintained the pristine sample's main shape after proton irradiation. In a study of the investigation of Raman peaks for ion (Mn+) bombarded ML-$MoS_2$ [40], the decrease in the intensities of the Raman peaks were attributed to the progressive ablation of the material caused by the ions bombarding the $MoS_2$ flakes. Hence, we can say no significant material removal detected through Raman spectroscopy. The ratio of the intensity of $A_{1g}$ to $E^1_{2g}$ modes (hereafter $R = \frac{I_{A_{1g}}}{I_{E^1_{2g}}}$) can be a measure of the total number of ions impinging on the surface ($1/L_D^2$ where $L_D$ is inter-defect distance)[26]. In the study of quantification of defects with Raman spectroscopy, Mignuzzi *et al.*[26] proposed a phenomenological relationship between the intensity ratio of the LA(M) peak and each of the first-order peaks, allowing a fast and practical quantification of defects in ML-MoS$_2$. They showed that $I_{LA}/I_{E^1_{2g}}$ and $I_{LA}/I_{A_{1g}}$ are inversely proportional to $L_D^2$, or, equivalently, directly proportional to the total number of ions impinging on the surface ($1/L_D^2$)[26]. They quantified the degree of the disorder by $L_D$. Furthermore, they showed that the intensity ratio of the LA(M) peak and each of the first-order peaks behave analogously to the intensity ratio of the D and the G peak, I (D)/I (G), in graphene, which is primarily used to assess the level of disorder within the lattice.[41] In our experiments, we did not observe the appearance of the LA(M) peak after irradiation. Hence, we cannot make a direct correlation between the degree of disorder and $L_D$. Based on our observation, while 150 keV irradiation only enhances the intensity ratio R by ~1% compared to the pristine sample, ~4% increase of the R-value was observed by 590 keV irradiation showing the enhanced interaction of electrons with $A_{1g}$ phonons. $L_D$ estimated by TEM measurements. $L_D$ increased with higher energy proton irradiation due to the induced changes in electronic band structure[42]. The $I_{A_{1g}}/I_{E^1_{2g}}$ ratio increases with increasing $L_D$, as the vacancy defects in the 590 keV irradiated sample decrease in density compared to 150 keV proton irradiated sample. For graphene, the ratio of I (D)/I (G) generally, increases with a decrease in defect densities[41]. We observed a similar effect in this case. The increase in the intensity of ratio R with increasing proton energy (where defect density decreases) needs to be further investigated, considering how radiation-induced defects modulate recombination centers' origin. We note that there can be slight differences between the frequencies of Raman lines in various studies due to differences in calibration/equipment[43], temperature[44], pressure[45], crystal size[46], all of which are possible expiations for these variations in reported wavenumbers. Hence, we can assume that proton irradiation did not induce significant changes in the electronic band structure of MoS$_2$ samples[42]. Kim *et al.*[16] showed that slight changes observed in Raman spectra after proton irradiation did not affect the electronic properties of MoS$_2$ devices. If researchers observe changes in electronic transport properties of devices made from these structures that would result from the charge traps in SiO$_2$ and at the MoS$_2$-SiO$_2$ interface, rather than by severe damage to the MoS$_2$ crystal structure. [16,47]

Table 1. The extracted parameters from Raman spectra for pristine and proton-irradiated samples

| Sample Ion Energy | Peak Position (cm$^{-1}$) | | Intensity | | FWHM (cm$^{-1}$) | | $R = \frac{I_{A_{1g}}}{I_{E^1_{2g}}}$ | $\Delta k = E^1_{2g} - A_{1g}$ (cm$^{-1}$) | Biaxial strain (%) | Uniaxial Strain (%) |
|---|---|---|---|---|---|---|---|---|---|---|
| | $E^1_{2g}$ mode | $A_{1g}$ mode | $E^1_{2g}$ mode | $A_{1g}$ mode | $E^1_{2g}$ mode | $A_{1g}$ mode | | | | |
| Pristine | 383.85 | 405.20 | 276.06 | 531.28 | 4.43 | 5.99 | 1.93 | 21.35 | - | |
| 150keV | 383.85 | 405.19 | 382.36 | 741.74 | 4.63 | 6.25 | 1.94 | 21.34 | None | None |

| 590keV | 383.49 | 404.80 | 336.99 | 675.49 | 4.48 | 5.67 | 2.00 | 21.31 | ~0.09 | ~0.2 |

The penetration depth distribution (stopping range) of protons, and the total vacancy concentration in the ML-MoS$_2$ were estimated using the Stopping and Range of Ions in Matter (SRIM)[48] simulator. For 150 and 590 keV protons irradiation, the penetration depth of proton beam was estimated to be 1.14 µm and 5.12 µm with a straggle (the spread of the ion beam energy spectrum) of 0.13 µm and 0.38 µm respectively, for stack MoS$_2$/SiO$_2$/Si (1.2 nm/300 nm/500 µm). Table 2 provides a summary of ion properties estimated by SRIM/TRIM.[48] The majority of protons could simply penetrate through the entire structure, while some amount of energy can also be transferred to the SiO$_2$ dielectric layer and monolayer, as predicted from the simulations. As shown by Kim *et al.*[16], this process creates electron-hole pairs and eventually induces charge traps in the oxide layer or at the interfaces. However, they did not observe any significant changes in the electrical parameters of the proton-irradiated devices with similar stacks of MoS$_2$/SiO$_2$/Si (2-8 nm/270 nm/500 µm) at a low fluence of $10^{12}$ proton/cm$^2$ as used in this experiment. Therefore we can say that low fluence proton irradiation effect is neither optical (as shown in the following paragraphs) nor electrical properties of MoS$_2$.

There are two types of interactions that result in a proton energy loss along the target material: (1) inelastic collisions with atomic electrons in the medium and (2) elastic collisions with nuclei of the atoms of the target material. While the ions' electronic energy loss helps recrystallization, nuclear energy loss is only responsible for defect creation[12]. The ratio between electronic and nuclear energy loss is 8% greater for a 590keV proton than a 150 keV proton irradiation. Observation of lower defect densities with higher energy irradiation in XPS and PL measurements might be related to this enhanced electronic energy loss component resulting in the recovery of the defective lattice despite the atomic displacements in the case of 590keV ion irradiation[49,40]. For the ranges calculated by SRIM, the electronic stopping dominates over the nuclear stopping throughout the structure.

Ionization (energy loss to electrons) values obtained from SRIM/TRIM simulations showed that energy loss to the target increases (28 eV/nm for 150 keV and 60 eV/nm for 590 keV) with the increase of the ion energy for the same fluence. The energy loss to electrons can be attributed to the charge transfer at the MoS$_2$/SiO$_2$ interface, which causes the XPS core-level peak shift, as suggested by Shi *et al.*[50] The fluence used here corresponds to a relatively low radiation dose compared to doses known to induce effects on the operation of TMD transistors due to anomalies between the material and oxide layers[1]. Furthermore, the energy loss to the target may produce vacancies. From our simulations, the density of vacancies was estimated to be in the order of 1.3 x $10^{16}$ cm$^{-3}$ and 2.1 x $10^{16}$ cm$^{-3}$ at the surface of the Si layers respectively for 150 keV and 590 keV proton fluences of $10^{12}$ proton/cm$^2$. Based on the information, we can say that there is no significant/detectable build-up of ions or damage in the ML- MoS$_2$ or in the dielectric due to proton irradiation, as confirmed by our Raman, PL, and TEM results. The number of vacancies depends on the displacement energy assigned to each target atom. It is assumed that the energy transferred into target phonons (10 eV/nm for 150 keV and 12 eV/nm) and the atom remains in its lattice site when the transferred energy is less than the displacement energy. The energy deposited to the target can be obtained by calculating the path length between two successive collisions and multiplying the specific energy loss[48]. As estimated by the SRIM program, energy transferred to the target is less than the displacement energies (Mo and S~25eV, Si~15eV, O~28eV)[48] of atoms in our stack. These results are consistent with the Raman spectroscopy and TEM results. The slight shift of the Raman frequencies can be referred to as energy is transferred into target phonons due to proton irradiation.

Table 2. Ranges of protons ions, longitudinal struggle, ionization, the density of defects, stopping power of irradiated $MoS_2/SiO_2/Si$ stuck at different energies Estimated from SRIM/TRIM

| Ion energy | Ion Range(μm) | Longitudinal straggle (A°) | Ionization (eV/nm) | Phonons (eV/nm) | Density of defects in Si (cm$^{-3}$) | (dE/dx)$_{elec}$ in $MoS_2/SiO_2/Si$ (keV/nm) | (dE/dx)$_{nuc}$ in $MoS_2/SiO_2/Si$ (keV/nm) |
|---|---|---|---|---|---|---|---|
| 150 keV | 1.14 | 1318 | 28 | 10 | $1.3 \times 10^{16}$ | $1.22 \times 10^{-1}$ | $1.66 \times 10^{-4}$ |
| 590 kev | 5.12 | 3777 | 60 | 12 | $2.1 \times 10^{16}$ | $6.25 \times 10^{-2}$ | $5.47 \times 10^{-5}$ |

Displacement threshold energy of 25 eV for Mo and S, 28 eV for O, and 15 eV for Si were used as suggested by SRIM results.

The X-ray photoelectron spectroscopy (XPS), a surface characterization technique, results indicate that a sample with the same growth conditions as the pristine sample is composed of Mo and S at an average Mo/S molar ratio of ~1:1.92, corresponding to $MoS_2$ with ~4% of sulfur vacancies, and no Na is detected from the basal plane of MoS2 flakes[24]. These measurements have been performed on another sample that has been grown at the same conditions to avoid ion exposure before proton irradiation. The pristine spectrum of S consists of S 2p3/2 and 2p1/2 peaks at 162.3 eV and 163.5 eV. The peaks at 229.5 eV and 232.6 eV observed in the pristine spectrum of Mo are identified as Mo 3d5/2 and 3d3/2. XPS results of the pristine sample are shown in Li *et al.*[24] After irradiation, XPS was used to characterize the sample after all the optical measurements were completed to avoid any possible damage due to ions. Our goal was to observe near-surface material modifications in atomic bonding and electronic structure that may change the transport properties of these materials if used in space electronics. The overall XPS spectrum measured for the irradiated $MoS_2$ showed carbon, molybdenum, sulfur, and oxygen. The concentration of Elements Detected (in atom %) were tabulated in Table 3

Table 3: The relative atomic concentration (in atom%) in the specimens' surface before and after irradiation.

| Sample | C | O | S | Si | Mo | S/Mo | S vacancy |
|---|---|---|---|---|---|---|---|
| Pristine | 14.7 | 40.6 | 9.46 | 21.8 | 4.73 | 1.92 | 4% |
| 590keV | 14.5 | 48.4 | 4.6 | 29.6 | 2.8 | 1.64 | 18% |
| 150keV | 22.6 | 40.0 | 7.2 | 25.1 | 5.1 | 1.41 | 30% |

The presence of oxygen and carbon on the pristine sample's surface can be due to the adsorption of common atmospheric gases during the growth or transferring process between the growth and measurements. The spectra include C1s peak approximately 284.08 and 284.29 eV for 150 keV and 590 keV proton irradiation energies, respectively. O1s peaks specifically O-Si peaks were at around 531.96 (for 150 keV) and 532.44 eV (for 590 kev) and 0-M peaks were at around 529.48 and 530.08, respectively, for 150keV and 590keV. No significant shifts in the peak position of C1s peak (~0.18 eV) were observed with higher energy irradiation. O1s peaks are upshifted 0.48 eV and 0.6 eV, respectively for O-Si and O-M peaks for 580 keV proton irradiation. The intensity of both C1s and O1s peaks is increased with the increased energy of proton irradiation. It was found that the surface concentrations of the C element decreased 55% with 590 keV proton irradiation as opposed to 150 keV irradiation while it stayed the same as compared to the pristine sample. For the O element, concentration increased 21 % with the 590 keV proton irradiation, as it stayed the same with 150 keV irradiation. These results indicate that some bonds, such as C–O, are broken during irradiation. The binding energies of Si 2p peaks are upshifted ~0.5 eV, and the concentration of Si elements is increased by 20% for higher energy proton irradiation. We would like to note that a shift of 0.2–0.3 eV might be insignificant since the XPS has a similar resolution. For both 150 and 590 keV proton irradiated $MoS_2$, Mo 3d spectra consist of peaks around 229.1 and 232.2 eV corresponding to the Mo $3d_{5/2}$ and Mo $3d_{3/2}$ components, respectively. An apparent sulfur 2s peak was observed at 226.3 eV for both 150 and 590 keV proton irradiations. Both 150 and 590 keV irradiated sample spectrum of S 2p3/2 peaks are observed at 161.9 eV. S 2p1/2 peaks were detected at 162.9 and 163.1 eV, respectively, for 150 and 590 keV proton irradiation. Mathew et al.[12] observed a similar peak position after 2MeV proton irradiation. There has not been any change in the position of Mo 3d and S 2p peaks that depend on proton irradiation energy. The binding energy values for Mo3d and S 2p peaks for proton irradiated samples shift downwards slightly ~0.4 eV compared to the pristine sample's binding energies regardless of the irradiation energy. This slight downshift of the binding energy can be attributed to the oxidation of the sample[51]. 150 keV and 590 keV protons penetrate 1.1 and 5.1 µm respectively deep into the Si substrate for stacks of $MoS_2$/$SiO_2$/Si (1.2 nm/300 nm/500 µm). Therefore, it is more likely that the observed peak shifts are due to proton-substrate interactions. Both Mo 3d and S 2p peaks at 590 keV radiated samples had increased by ~50% in intensity compared to that in the 150 keV irradiated sample. At around 232.7 eV and 235.5 eV (235.8 for 590 keV) possible $MoO_3$ peaks were noticed for 150 keV irradiated samples. The observed $MoO_3$ peaks might result from the oxidative nature of protons due to the exposure to the atmosphere during measurements. On the other hand, it is usual to observe oxidized molybdenum in $MoS_2$ samples[52,53], and these can be removed after annealing of the samples[54]. We would like to note that $MoO_3$ is hard to resolve from $MoS_2$ plasmon loss peaks in the Mo 3d spectra. Therefore, it is less likely that the slight changes in binding energy are due to valence states which are induced by oxidation. The downward shift of binding energies may also be attributed to the composition deviation from the stoichiometry at the $MoS_2$ surface[51]. Baker et al.[54] showed that the binding energy of Mo-3d5/2 decreased progressively as the x in $MoS_x$ decreased. The Mo/S molar ratio has increased to 1:1.41 with radiation of 150keV, while a small increase of the ratio 1:1.64 was observed for 590 keV proton irradiation. This indicates a 26% and 14% increase in the sulfur vacancies respectively for 150 keV and 590 keV irradiation. Higher S vacancies with lower energies might be because the penetration of the proton is lower, and particles stop closer to $MoS_2$ flakes. Liu

*et al.*[19] have also observed a larger content ratio of Mo/S after irradiation. The S/Mo molar ratio is <2.0 elucidates that the n-type nature of MoS₂ since the locally formed sulfur vacancies act as electron donors[51]. In this study, we displayed that the n-type nature of MoS₂ increases with the decrease of proton energies. This can also indicate that the Fermi-level of ML-MoS₂ shifted upwards after proton irradiation. Kwon *et al.*[11] showed that the n-type doping effect of MoS₂ with intrinsic n-type semiconducting property increases with an increasing number of protons irradiated on ML-MoS₂. They found that work function decreases, and contact potential difference increases with the increase of S vacancies due to proton irradiation. This information might be useful for device fabrication in determining the properties of gold contact and modifying the electrical properties of ML-MoS₂.

Both MoS₂ and the S 2p peaks were broader with the 150 keV irradiation than 590 keV proton irradiation. These results align with the PL spectrum results, where slightly higher defect concentration was observed with lower energy proton irradiation. Significant shifts in peak positions for carbon (C 1 s) or silicon (Si 2p) were not detected after irradiation. In summary, increase defect density observed by the PL spectrum can be attributed to sulfur vacancies. However, these vacancies are not adequate to cause substantial changes in the Raman spectroscopy or change electronic band structure.

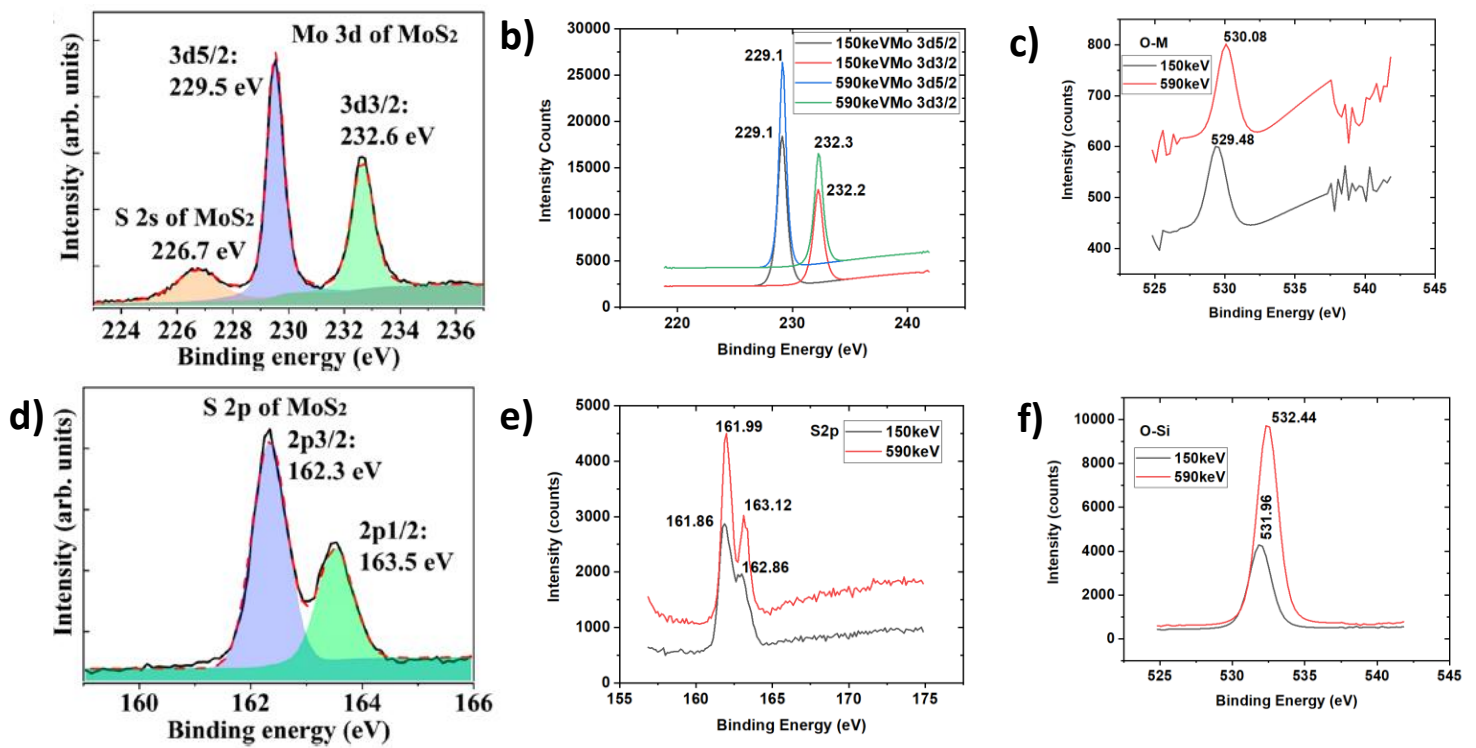

**Figure 3: XPS spectra from pristine and irradiated MoS2 at a fluence of 1x10¹²ions/cm².** Species detected included: $MoS_2$, hydrocarbons, $SiO_2$, and likely $MoO_3$. The Mo peak is given in (a) for pristine and (b) for radiated and S peak in (d) for pristine and (e) radiated. Oxygen binding to Mo (c) and Si (f) after irradiation. The fitted spectra, along with the constituent peaks and experimental points, are also shown.

Figure 4 shows the optical morphology of the samples obtained by atomic force microscopy (AFM) before and after proton irradiation. The AFM images of the pristine MoS$_2$ represent the flat terraces and no distinctive features shown in another article[24]. The region of the relatively light color surface exhibits the ML-MoS$_2$ while the darker surface region indicates the substrate. The height difference between the substrate and monolayer did not change with proton irradiation, indicating the maintenance of monolayer futures of MoS$_2$. The bright point-like features (Fig. 4c) between the substrate and the monolayers are NaBr nanoparticles, as stated before by Li *et al.*[24] AFM images of ML- MoS$_2$ exposed to the lowest energy show many randomly distributed hillocks (see Figures 4a and 4c), which are not present on pristine samples. The 590 keV proton irradiated ML-MoS$_2$ is easier to distinguish from the substrate than the one irradiated with 150 keV proton since there were many small particulates. These small hillocks could be the remnants of the ejection of sulfur from the surface after irradiation, as suggested by the increase of the sulfur vacancies in the XPS measurements. The apparent height (~2.5-5 nm) of these protrusions depends strongly on the scanning parameters, and they can be seen even more clearly in the corresponding phase images.

The sample's average roughness increased from 0.35 nm to 0.48 nm with the decrease of the proton irradiation energy. These results align with the finding of XPS and PL, where higher defect densities observed with lower energy proton irradiation. On the other hand, the height difference between the MoS$_2$ flake and the substrate of ~1.2 nm did not change with different radiation energies. Previously Kwon *et al.*[11] did not observe any damages on the surface of ML-MoS$_2$ for 10MeV proton irradiation at the same flux.

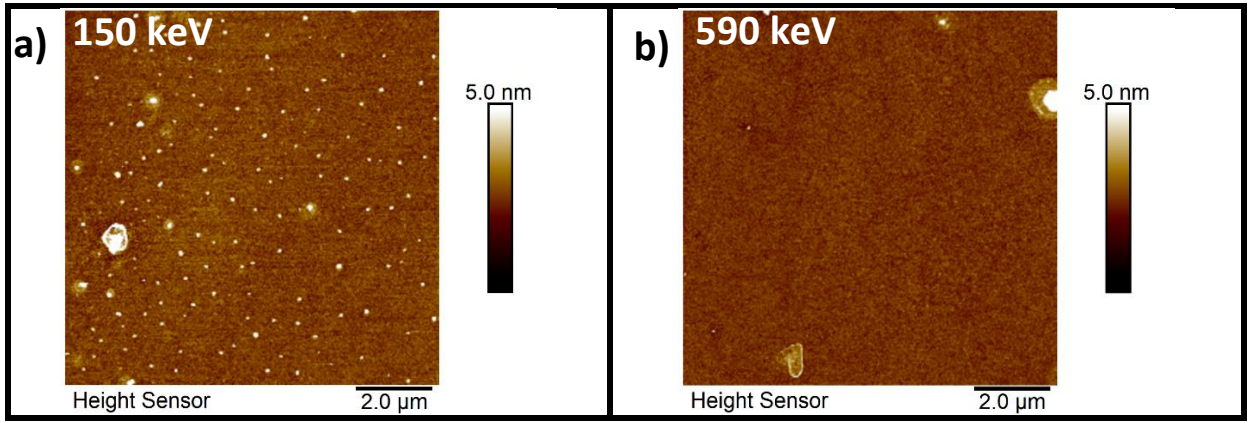

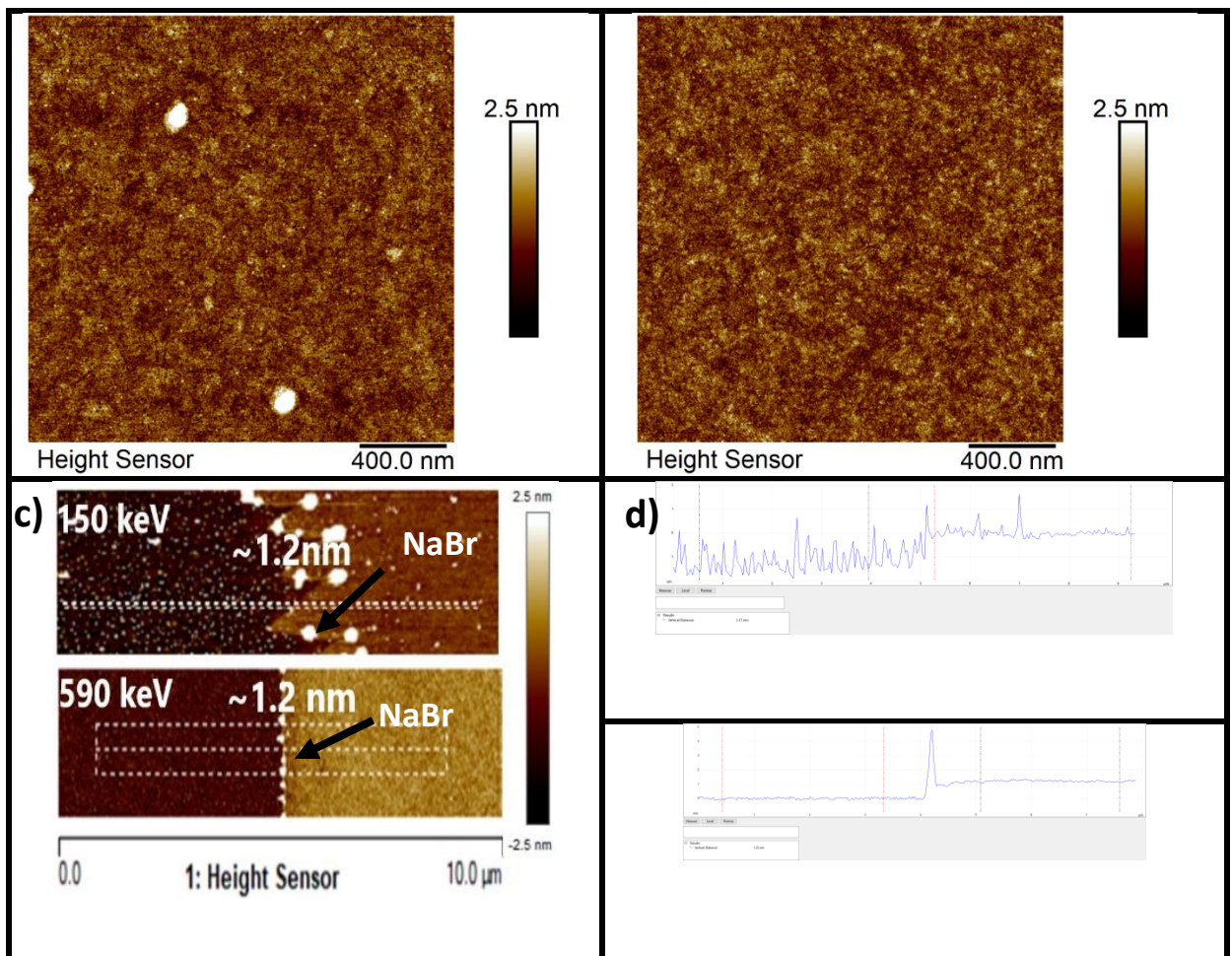

**Figure 4: AFM micrograph of the proton irradiated MoS$_2$**: (a) Protrusions and small holes were observed on the surface after 150 keV irradiation (b) 590 keV irradiated sample. The root mean square value of surface roughness (Rq) was 0.48 nm and 0.35 nm for the 150 keV and 590 keV proton-irradiated samples (c)AFM image of a single-crystal ML-MoS$_2$ with NaBr nanoparticles (as indicated by the arrow) on the edges (d)AFM height profile along the arrowed dashed white boxes in confirms the monolayer step height and shows the several nanometer-sized NaBr particles on edge.

Scanning electron microscopy (SEM) can cause contamination and potential quenching of the MoS$_2$; hence images were taken after all optical measurements. We also repeated optical measurements after SEM measurements yet we did not observe any changes. The SEM surface morphologies of MoS$_2$ before and after proton irradiation are shown in Figure 1. The sample was undamaged before irradiation. However, we observed the existence of some grain boundaries after irradiation. The degree of erosion increased with increasing proton irradiation energy. The existence of grain boundaries with white particles on them can be explained through the complex chemical and physical surface material reactions resulting in volatile products and then removed from the surface as suggested by Liu *et al.*[19] Finally these reactions cause changes in the surface morphology agrees with the decreasing trend of roughness with higher energy irradiation as shown in the AFM results.

TEM was applied to identify the possible defect sites introduced after irradiation. Figure 5 shows the atomic resolution images of irradiated $MoS_2$ samples transferred onto Quantaifoil TEM grids. TEM image shows alternating bright Mo and less bright S sites arranging in hexagonal rings. While the brighter lattice sites correspond to Mo, $S_2$ lattice sites appear to be darker in this image due to Z-contrast (ZMo = 42, 2 × ZS = 32)[21]. Relatively light grey regions are surface contaminations due to the transfer of the sample to the TEM grid. Extra bright white dots might be either surface contaminations or free atoms since they move around when we move the TEM beam. A plot of intensity versus distance for a selected profile represented by the red arrow in Fig. 5(b), shows a projected "in-plane" $S_2$–Mo lattice site lateral distance of (to be calculated) Å. Low spatial frequency intensity fluctuations in the TEM image is attributed to residual surface contamination[21]. Fig. 5(b and c) shows some suspected defect sites (mainly S vacancies) observed after irradiation and are circled in the TEM image of irradiated $MoS_2$. While it is challanging to identify sulfur vacancies in the TEM image of 590 keV irradiated sample, relatively more suspected S vacancies are observed on the 150 keV irradiated samples and highlighted in Fig. 5. These results align with the sulfur vacancy increases observed with proton irradiation via XPS measurements. It is common to observe the point defects such as vacancies in CVD grown MoS2 because of the growth process[55]. Therefore, it is expected to observe some preexisting vacancies even in irradiated samples. The average inter-defect distance ($L_D$) is inversely proportional to the square root of the ion dose (σ) density[26]. The proton fluence of $10^{12}$ proton/cm² corresponds to an average $L_D$ of 10 nm ($L_D = 1/\sqrt{\sigma}$). These distances are estimated to be changing between 5 and 10 nm (see Figure 5 b & c). The high energy of hydrogen ions creates gentle changes in the crystallographic structure of ML $MoS_2$ compared to results with low energy protons.

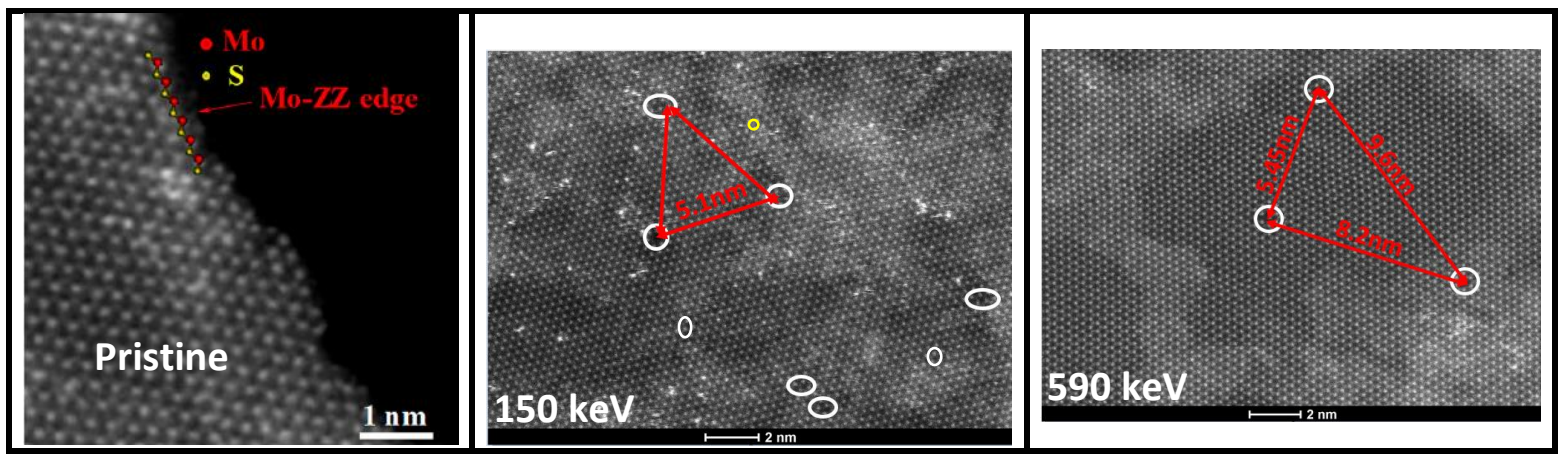

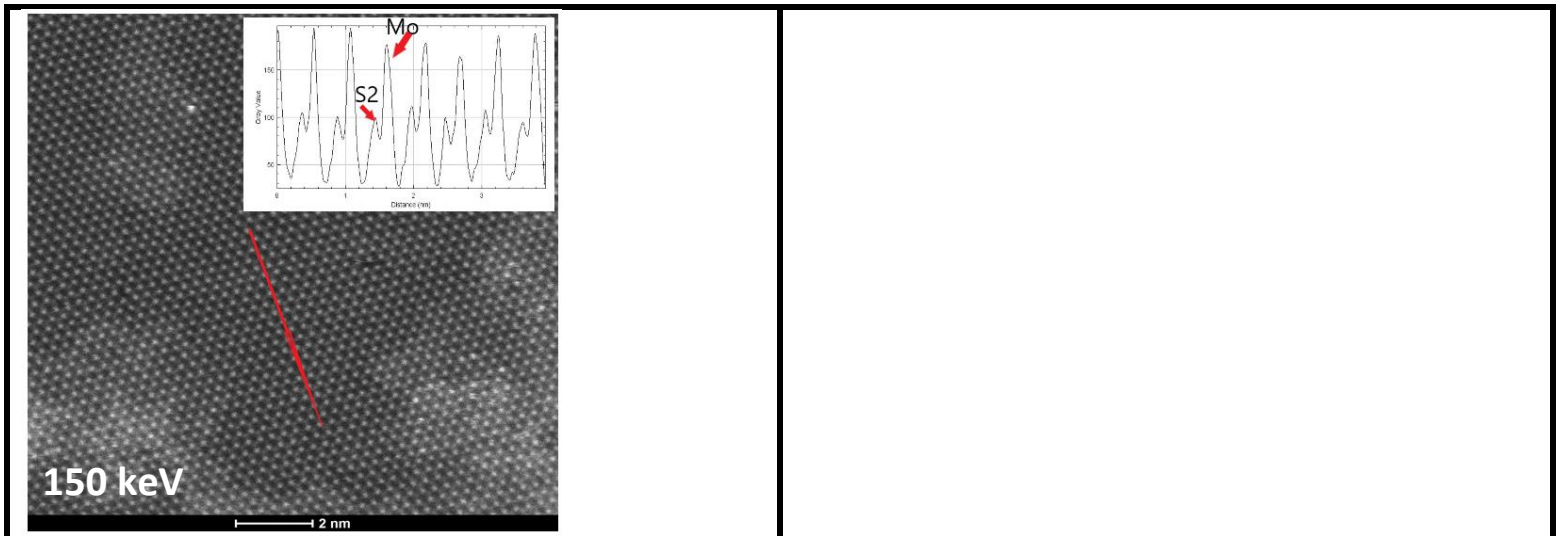

**Figure 5: TEM images of the pristine and irradiated samples** (a) pristine MoS$_2$ region resolving individual Mo and S$_2$ crystal lattice sites. The image data used here is taken from another study[24]. (b) The image data from 150keV irradiated MoS$_2$. (c) The image data from 590 keV irradiated MoS$_2$. Suspected S vacancy defect sites are circled. Possible reasons can be that the S atoms are hard to identify in STEM images due to low contrast since the image is not filtered, or those dangling S atoms have been reconstructed under electron beam irradiation or removed during the transfer process for TEM as suggested by Li *et al.* [24] (d) Image date show a projected "in-plane" S2–Mo lattice site lateral distance for irradiated samples.

## CONCLUSION

In conclusion, we have studied the effects of low energy proton irradiation on Raman, PL, and XPS spectra from ML-MoS$_2$ flakes grown with the facilitation of sodium bromide by the chemical vapor deposition and its morphological characteristics imaged with SEM, TEM, and AFM imaging in the vision of applications in space exploration We have demonstrated that radiation damage to the surface of the monolayer and surrounding materials does not influence the optical properties of the MoS$_2$ material at low fluences significantly. This result is also consistent with previous studies[16,17]. The relatively low damage resulting from these low radiation energies can be attributed to the atomically thin nature of MoS$_2$. Our results provide useful insights into the radiation-induced morphological, chemical, and optical changes that can be beneficial during the fabrication of devices, which are resilient to radiation damages from energetic charged particles. Finally, our results can be an indication of MoS$_2$ being an excellent candidate for electronic materials in future space applications and other high-radiation environments in LEO. Future study of electrical transport measurements of MoS$_2$ devices before and after irradiation, and time-resolved PL measurements might be useful for further quantifying the radiation-induced effects while observing their effect on device performance.

## Acknowledgment

The authors would like to acknowledge the financial support from Penn State Abington Faculty Development Grant. The authors thank Mr. Wesley Ellis Auker for SEM measurements, Dr. Timothy B Tighe for AFM measurements, Mr. Jeffrey Shallenberger for XPS measurements, and Mr. Max Cichon for proton irradiation. In addition, the authors would like to acknowledge Dr. Xufan Li for valuable discussions.